\documentclass[12pt]{article}
\usepackage{graphicx}
\usepackage{epsfig}
\usepackage{amsbsy}

\begin{document}
\baselineskip 10mm

\centerline{\bf Structure and Stability of Two-Dimensional Complexes
of C$_{20}$ Fullerenes}

\vskip 4mm

\centerline{I. V. Davydov, A. I. Podlivaev, and L. A. Openov$^{1)}$}

\vskip 4mm

\centerline{\it Moscow Engineering Physics Institute (State University)} \centerline{\it Kashirskoe sh. 31, Moscow, 115409 Russia}

\vskip 15mm

$^{1)}$e-mail: LAOpenov@mephi.ru

\vskip 15mm

\centerline{\bf ABSTRACT}

Two-dimensional systems of $\mathrm{C}_{20}$ fullerenes connected to each other by strong covalent bonds have been investigated.
Several isomers differing in the type of intercluster bonds have been revealed. The lifetimes $\tau$ of the $(\mathrm{C}_{20})_{\textnormal{M}\times\textnormal{M}}$ complexes with $M=$ 2 and 3 at $T\,=\,(1800\,-\,3300)$ Ê have been directly calculated using the molecular dynamics method. It has been shown that these complexes lose their periodic cluster structure due usually to the coalescence of two or several neighboring $\mathrm{C}_{20}$ fullerenes. The activation energy of this process determined by analyzing the $\tau\left(T\right)$ dependence appears to be $E_{a}\approx2.5$ eV in agreement with the calculations of the heights
of the potential barriers preventing the coalescence. At high temperatures $T>2400$ K, the decay of $\mathrm{C}_{20}$ fullerenes entering into the complex is possible.

\vskip 15mm

PACS: 36.40.Qv,61.48+c,71.15Pd

\newpage

The minimum possible fullerene $\mathrm{C}_{20}$ has attracted
considerable interest since its discovery in 2000 [1]. Great attention is focused on the study of the possibility of existing macroscopic crystalline forms based on $\mathrm{C}_{20}$ fullerenes similar to $\mathrm{C}_{60}$ fullerites [2,3]. According to theoretical works [4-7], $\mathrm{C}_{20}$ clusters in crystals are connected with each other by strong covalent bonds. This property distinguishes the $\mathrm{C}_{20}$ fullerite from the $\mathrm{C}_{60}$ fullerite, where the bond between $\mathrm{C}_{60}$ clusters is due to a weak van der Waals attraction. Interest in the $\mathrm{C}_{20}$ fullerite is stimulated in particular by the expectation that the superconducting transition temperature
in it will be higher than that in the $\mathrm{C}_{60}$ fullerite
(significant enhancement of the electron-phonon interaction
with a decrease in the size of fullerites from
$\mathrm{C}_{60}$ to $\mathrm{C}_{36}$, $\mathrm{C}_{28}$, and
$\mathrm{C}_{20}$ was predicted [8]).

There is experimental evidence for the existence of
$(\mathrm{C}_{20})_{\textnormal{N}}^{+}$ complexes with $N=3\div13$ [9]
(which are likely chains of $\mathrm{C}_{20}$ fullerenes [10, 11]) and bulk phases [12,13] in which $\mathrm{C}_{20}$ fullerenes form a crystal lattice as atoms in normal solids. The structure of these phases is still unknown (the transmission electron microscopy data [12] are inconsistent with the electron diffraction data [13], indicating that the phase observed in [12] differs from that observed in [13]). The density functional theory investigations [4,5,7] show that it can be both simple cubic and body-centered (cubic, tetragonal, or orthorhombic). For the same lattice type, the intercluster bonds can have various forms [7], leading to a large
variety of atomic configurations, candidates for the $\mathrm{C}_{20}$ fullerite. All of these configurations differ in the binding energy of $\mathrm{C}_{20}$ fullerenes in a crystal, $\Delta
E=E\left(\mathnormal{\textnormal{C}_{20}}\right)-E\left[\left(\mathnormal{\textnormal{C}_{20}}\right)_{\textnormal{N}}\right]/N$, where $N$ is the number of fullerenes ($N\gg1$ in the macroscopic system). The configuration with the maximum $\Delta E$ value is the most energetically favorable one (it has the minimum total energy) and the remaining configurations are metastable with respect to this configuration. Moreover, as shown in [13], a crystal of $\mathrm{C}_{20}$ fullerenes can have the fcc lattice, where
$\mathrm{C}_{20}$ fullerenes are connected with each other by means of
additional carbon atoms, so that the unit cell consists of 22 atoms. Other three-dimensional periodic configurations, which are obtained from various combinations of $\mathrm{C}_{20}$, likely exist (the high chemical activity of $\mathrm{C}_{20}$ fullerenes can lead to the formation of amorphous
structures in them).

Note that even in the simplest $(\mathrm{C}_{20})_{\textnormal{N}}$ systems, quasi-one-dimensional chains, the number of various types of intercluster bonds is quite large [7,10,11]. In these systems, the $\Delta E$ value is maximal in chains with the \textit{open}-{[}2+2{]} bonds and is somewhat smaller in chains with {}``twisted'' (according to [7]) bonds. Before the search for the three-dimensional $(\mathrm{C}_{20})_{\textnormal{N}}$ structure with the maximum $\Delta E$ value, it is reasonable to study in detail two-dimensional complexes of $\mathrm{C}_{20}$ fullerenes. Although the data on the synthesis of cluster films of $\mathrm{C}_{n}$ with $n=20-32$ exist [14] (but have not yet been confirmed), this problem is poorly investigated theoretically [15].

Since the strong interaction between clusters in $\mathrm{C}_{20}$ fullerite can give rise to the transition to another atomic configuration (not even crystalline, but amorphous), in addition to the energy benefit of a certain structure, its stability is important. As shown in our previous work
[16], an isolated $\mathrm{C}_{20}$  {} fullerene loses its spheroidal shape through the decay into various energetically unfavorable
configurations, rather than through its transition to the equilibrium ``bowl'' isomer with a minimum total energy. The chain-cluster structure of quasi-one-dimensional $(\mathrm{C}_{20})_{\textnormal{N}}$ complexes can disappear due both to the decay of one of the $\mathrm{C}_{20}$ fullerenes and to the coalescence of two neighboring $\mathrm{C}_{20}$ into a $\mathrm{C}_{40}$ cluster (the separation of one of the fullerenes from the chain is also possible) [10, 11]. In this case, the height $U$ of the lowest
potential barrier around the corresponding metastable configuration decreases from $\approx$ 5 eV in the $\mathrm{C}_{20}$ fullerene [16] to $\approx$ 2.5 eV in the $(\mathrm{C}_{20})_{2}$ dimer [17] and further to $\approx$ 2 eV in the $(\mathrm{C}_{20})_{\textnormal{N}}$ chains with $N\gg1$, where the main stability loss channel is the coalescence of
fullerenes [10,11]. In two-dimensional systems, both the stabilization of the metastable state due to an increase in the number of the nearest neighbors of each fullerene and the appearance of new stability loss
mechanisms can be expected.

The aim of this work is to analyze the structure and energy characteristics of two-dimensional complexes of $\mathrm{C}_{20}$ fullerenes, as well as their stability. The tight-binding method [18], which includes the contribution of all valence electrons (four per atom) to the total
potential energy $E$, is used in the calculations. This method does not require such computational resources as ab initio approaches and allows the investigation of systems consisting of several hundreds or more atoms.
For $\mathrm{C}_{60}$ and $\mathrm{C}_{20}$ clusters, it provides the interatomic distances, binding energy, HOMO-LUMO gap, and other
characteristics most of which are in good agreement with the experimental data and ab initio calculations [16,17,19-21]. For the
$(\mathrm{C}_{20})_{\textnormal{N}}$ chains, it gives the same
sequence in the location of isomers with various intercluster
bonds in the $\Delta E$ value [10,11] as more strict density-functional calculations [7].

Making use of free boundary conditions, we investigate quasi-two-dimensional
$(\mathrm{C}_{20})_{\textnormal{M}\times\textnormal{M}}$ complexes with
$M=2-5$ and various types of intercluster bonds. By analogy
with the $(\mathrm{C}_{20})_{\textnormal{N}}$ chains [10,11], it can be expected that the binding energy $\Delta E$ is maximal when $\mathrm{C}_{20}$ fullerenes are connected with each other by \textit{open}-{[}2+2{]} bonds in
both horizontal and vertical directions (configuration A, Fig. 1a). However, the $\Delta E$ value appears to be much higher (by $\sim 1$ eV/$\mathrm{C}_{20}$) if intercluster bonds are twisted along one of the directions (configuration B, Fig. 1b). Configuration B is obtained from configuration A by the shift of the horizontal chains with respect to each
other so that the intercluster bonds in the vertical direction (i.e., bonds between fullerenes of different chains) become twisted. This configuration can be treated also as a configuration where vertical chains with twisted bonds are connected by the \textit{open}-{[}2+2{]} bonds. The difference between the binding energies $\Delta E$ in the
$(\mathrm{C}_{20})_{\textnormal{M}\times\textnormal{M}}$ complexes
with configurations A and B increases with the size of the system (i.e., with $M$). The extrapolation of the data obtained for $M=2-5$ to the limit
$M\rightarrow\infty$ yields $\Delta E=8.00$ eV/$\mathrm{C}_{20}$ and 9.54
eV/$\mathrm{C}_{20}$ for configurations A and B, respectively (see Fig. 2). We also calculated $\Delta E$ for configuration B with periodic boundary
conditions. The value $\Delta E=(9.45\div 9.81)$ eV/$\mathrm{C}_{20}$ obtained by extrapolating to the limit $M\rightarrow\infty$ almost does not differ from the value obtained with free boundary conditions (a comparatively large error is due to the
nonmonotonic saw-tooth dependence of $\Delta E$ on $M$ and
the small number of points used for extrapolation).

To analyze the thermal stability of the
$(\mathrm{C}_{20})_{\textnormal{M}\times\textnormal{M}}$ complexes, we used the molecular dynamics method. Configuration B with the combination of the \textit{open}-{[}2+2{]} bonds and twisted intercluster bonds (see Fig. 1b)
was taken as initial. At $t=0$, all atoms were assigned with random velocities and displacements so that the total momentum and total angular momentum were zero. Classical Newton's equations of motion
were solved numerically with a time step of $t_{0}=2.72\times10^{-16}$ s. The total energy of the system (the sum of the potential and kinetic energies) remains unchanged in the simulation process, corresponding to a microcanonical ensemble, and its dynamic temperature $T$, which is
a measure of the energy of the relative atomic motion, was calculated by the formula [22,23] $\left\langle
E_{\mathnormal{\textnormal{kin}}}\right\rangle
=\frac{1}{2}k_{\textnormal{B}}T\left(3m-6\right)$, where $\left\langle
E_{\mathnormal{\textnormal{kin}}}\right\rangle$ is the time-averaged
kinetic energy of the entire system, $m=20M^{2}$ is the
number of atoms in it, and $k_{\textnormal{B}}$ is the Boltzmann constant.
Since the lifetime of the $(\mathrm{C}_{20})_{\textnormal{M}\times\textnormal{M}}$ complexes to
the instant of loss of their periodic cluster structure increases exponentially with a decrease in $T$ and the time necessary for the numerical calculations at a given $T$ value is proportional to
$M^{6}$, we restricted ourselves to the systems with $M=2$ and 3 at
$T\,=\,(1800\,-\,3300)$ K.

Sometimes we observed transitions from configuration B to configuration A and vice versa. This is due to the nearly equal values of the energies of these configurations (see Fig. 2) and likely to a quite low potential barrier
between them (according to the results obtained for the $(\mathrm{C}_{20})_{\textnormal{N}}$ chains [10], its height is expected to be about 1 eV). However, although such transitions are rather fast
(occur in a time of 0.1-1 ps corresponding to $\sim 1000$ molecular dynamics steps), they are sufficiently rare and do not lead to the breakdown of the spheroidal shape of $\mathrm{C}_{20}$ fullerenes or to the breaking of the periodicity in their arrangement.

As in the simulation of the $(\mathrm{C}_{20})_{\textnormal{N}}$
chains [11], we observed two fundamentally different stability loss
channels for the$(\mathrm{C}_{20})_{\textnormal{M}\times\textnormal{M}}$: the decay of one of the $\mathrm{C}_{20}$ fullerenes and coalescence of two neighboring $\mathrm{C}_{20}$ fullerenes into the $\mathrm{C}_{40}$ cluster. The decay and coalescence are accompanied by a decrease and an increase in the binding energy $\Delta E$ by $\left(0.5\,-\,1\right)$ eV respectively;
i.e., the system transits to the energetically less favorable and more favorable configurations, respectively. In both cases, the first decay/coalescence act is followed by the decay and/or coalescence of the
remaining fullerenes of the complex (see Fig. 3). In particular, it is sometimes difficult to distinguish whether the sequential coalescence of two $\mathrm{C}_{20}$ fullerenes into the $\mathrm{C}_{40}$ cluster occurs followed by its coalescence with another $\mathrm{C}_{20}$ fullerene into the $\mathrm{C}_{60}$ cluster or three $\mathrm{C}_{20}$ fullerenes immediately fuse into the $\mathrm{C}_{60}$. cluster. For the same reason, it is sometimes impossible to clearly determine which event - decay or coalescence - occurs first initiating the integrity loss of the entire system. Nevertheless, the temperature of the
$(\mathrm{C}_{20})_{\textnormal{M}\times\textnormal{M}}$ complex correlates with the stability loss mechanism: at $1800<T<2400$ K, the periodic
cluster structure disappears due only to the coalescence of
$\mathrm{C}_{20}$ fullerenes; whereas at $T>2400$ K, due both to the coalescence of $\mathrm{C}_{20}$ fullerenes and to their
decay, although the decays occur much rarely (see Fig. 4). Approximately the same processes occur in the $(\mathrm{C}_{20})_{\textnormal{N}}$ chains [11]; however, in contrast to the chains, the separation of one $\mathrm{C}_{20}$ fullerene from the
$(\mathrm{C}_{20})_{\textnormal{M}\times\textnormal{M}}$ complex is not observed probably because each $\mathrm{C}_{20}$ fullerene has a covalent bond with no less than two other fullerenes.

To determine the activation energy $E_{a}$ for the coalescence
of the $\mathrm{C}_{20}$ fullerenes in the $(\mathrm{C}_{20})_{2\times2}$ complex, we analyzed the temperature dependence of the lifetime using the Arrhenius formula for the microcanonical ensemble with the finite-heat-bath correction [24,25]. We assumed that $\tau^{-1}\left(T\right)=A\exp\left(-E_{a}/k_{\mathnormal{\textnormal{B}}}T^{*}\right)$, where $T^{*}=T-E_{a}/2C$ (the microcanonical heat capacity is assumed to be $C=\left(3m-6\right)k_{\textnormal{B}}$, where $m=20M^{2}$; we will show below that $E_{a}/2C\approx100$K$\ll T$ at $T\approx$ 2000 K). As seen in Fig. 4, the calculated dependence of $\ln\left(\tau\right)$ on $1/T^{*}$ in the first approximation is given by a straight line (because of the rather large spread of the {}``numerical experiment'' data, the temperature dependence of the parameter $A$ cannot be reliably determined). Using the slope of this straight line and the point of its intersection with the ordinate axis, we determined the activation energy $E_{a}=\left(2.45\pm0.2\right)$ eV and the frequency factor
$A=\left(6\pm1\right)\times10^{17}\textnormal{s}^{-1}$. The $E_a$ value coincides within the errors with $E_{a}=\left(2.7\pm0.3\right)$ eV for the coalescence of $\mathrm{C}_{20}$ fullerenes into the $(\mathrm{C}_{20})_{2}$
dimer [21]. This value is generally averaged over the different coalescence
channels. Indeed, analysis of the potential energy of the $(\mathrm{C}_{20})_{2\times2}$ complex as a function of the coordinates
of constituent atoms shows that there are many different paths for the transition from configuration B to the configurations with the fusedfullerenes and that the heights $U$ of the potential barriers on these paths vary in a quite wide range of $U$ = (2.4 - 3.0) eV (we do not
exclude the possibility that barriers with a lower $U$ value exist, but we have not found them due to the complex shape of the potential relief).

In order to collect the simulation results sufficient for determining $E_{a}$ in the $(\mathrm{C}_{20})_{3\times3}$ complex, a very long
computer time is required. Nevertheless, the comparison of $\tau$ values obtained for certain temperatures in the range $T=\left(2000-3000\right)$ K with the $\tau\left(T\right)$ dependence for the coalescence of the $\mathrm{C}_{20}$ fullerenes in the $(\mathrm{C}_{20})_{2\times2}$ complex (see Fig. 4) shows that the stability of the system remains almost unchanged with an increase in the system size: $E_{a}\approx2.5\,$ eV for both $M=2$ and 3. The complete absence of the decay processes at $T<2400$ K
indicates that their activation energy is quite high (likely, $E_{a}=\left(3.5-4\right)\,$ eV, as for the
$(\mathrm{C}_{20})_{2}$ dimer [21]).

Summarizing the results, we can state that the main channel of the breakdown of the periodic cluster structure of the two-dimensional $(\mathrm{C}_{20})_{\textnormal{M}\times\textnormal{M}}$
systems is the coalescence of their constituent $\mathrm{C}_{20}$ fullerenes. The average activation energy of such processes is about 2.5 eV; i.e., the lifetime of these systems at room temperature is macroscopically long. Since the stability of the two-dimensional complexes of the $\mathrm{C}_{20}$ fullerenes is no worse than the stability of one-dimensional chains [11], it can be expected that the three-dimensional systems of the $\mathrm{C}_{20}$ clusters are also characterized by a quite high stability.

To conclude, we note that the density of the single-particle electronic states at the Fermi level in two-dimensional $(\mathrm{C}_{20})_{\textnormal{M}\times\textnormal{M}}$ complexes is finite (see Fig. 5) and depends weakly on $M$ for $M=3-5$. For this reason, it can be expected that, if films of $\mathrm{C}_{20}$ fullerenes were synthesized, they will be conducting. To manufacture such films, one can try to modify the method of low-energy beam deposition of neutral clusters [14]. In this case, it is very important to select the appropriate substrate material (e.g., $\mathrm{C}_{60}$ monolayers were recently grown on Au substrates [26]). It is also of interest to perform the numerical simulation of the coalescence of initially isolated $\mathrm{C}_{20}$ fullerenes into two-dimensional complexes.

\newpage
\includegraphics[width=\hsize,height=15cm]{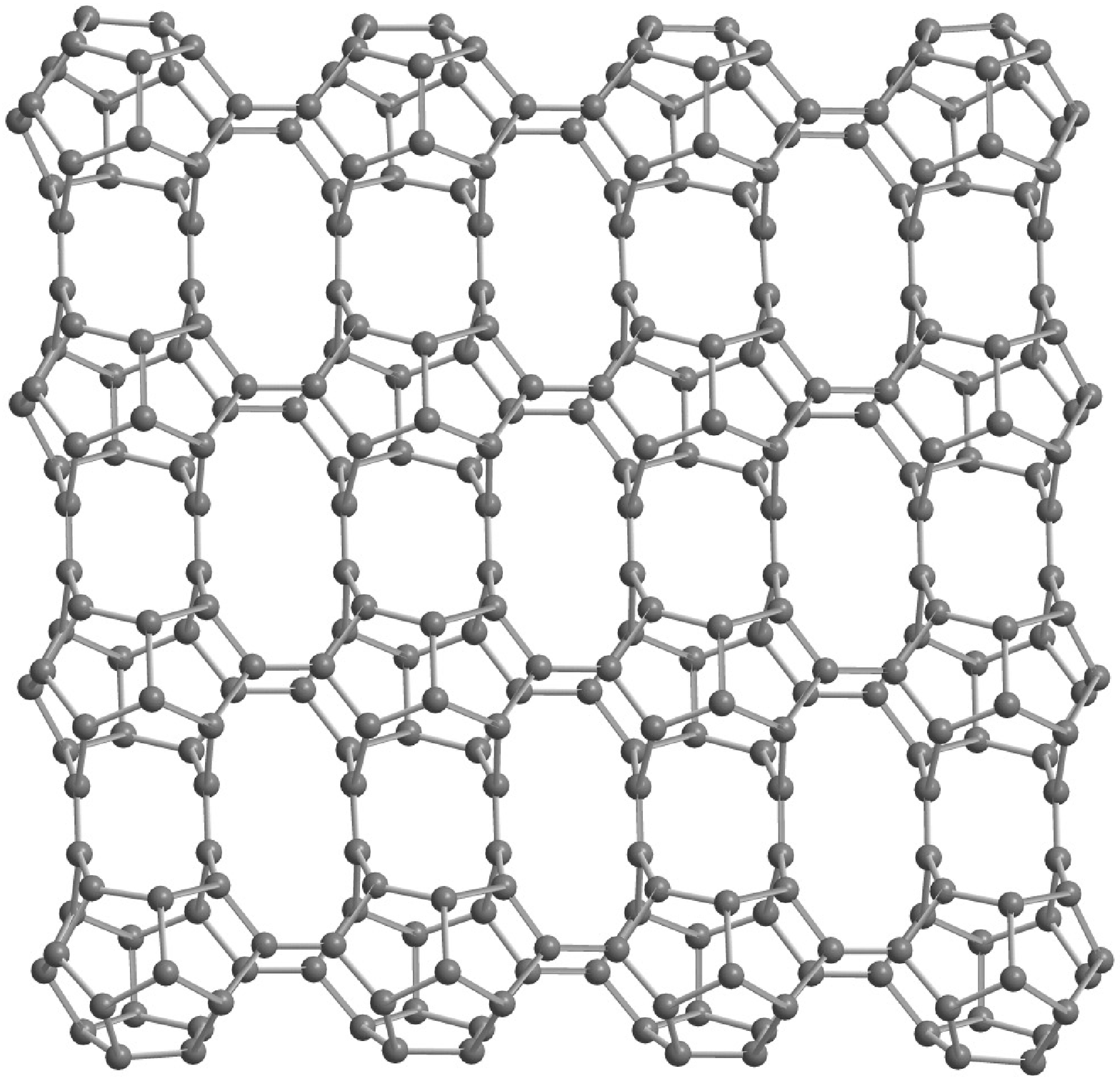}
\vskip 15mm
Fig.1a. Two-dimensional $(\mathrm{C}_{20})_{4\times4}$ structures with the \textit{open}-{[}2+2{]} intercluster bonds, configuration A. The binding energy of the fullerenes is $\Delta E=6.11$ eV/$\mathrm{C}_{20}$.

\newpage
\includegraphics[width=\hsize,height=15cm]{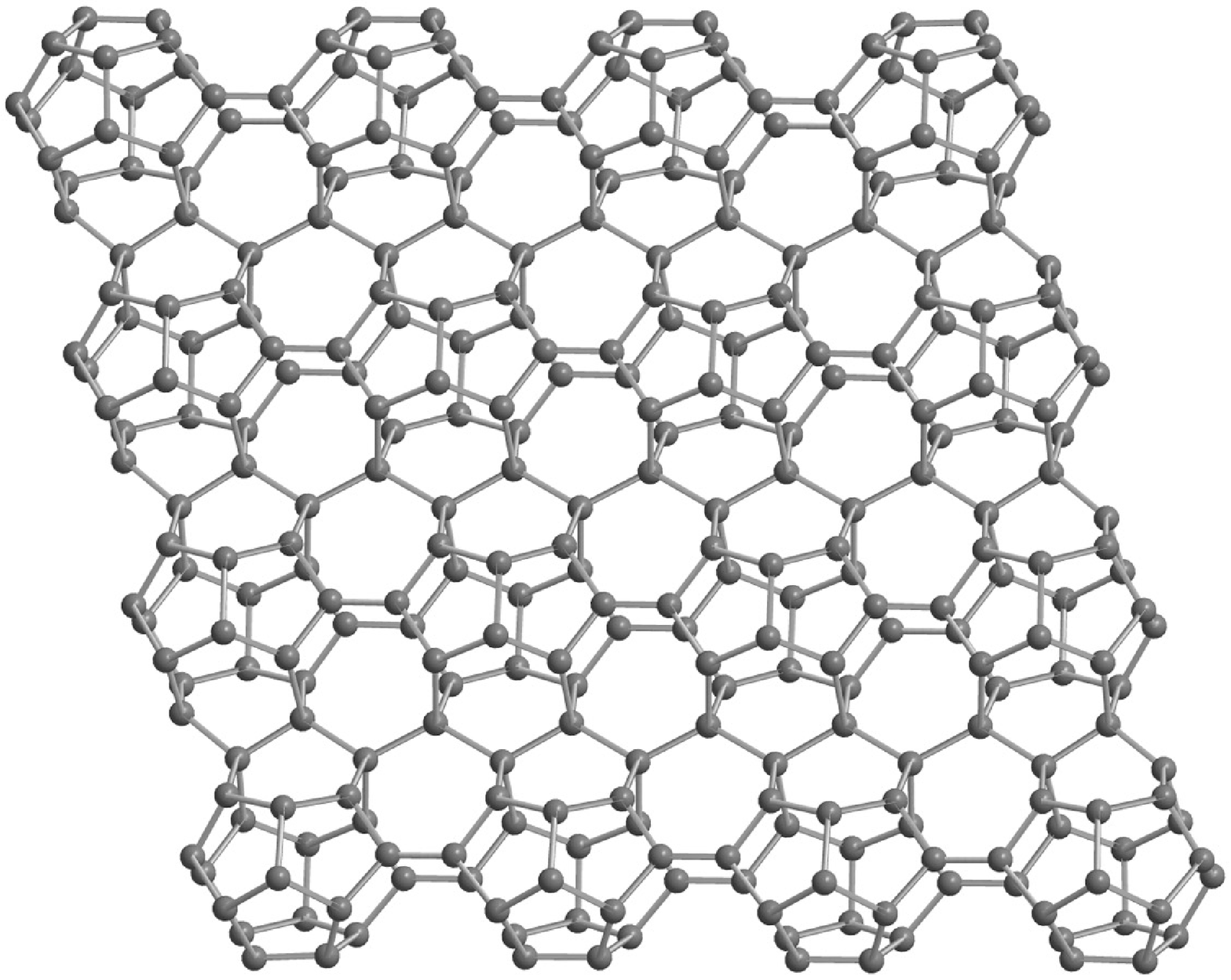}
\vskip 15mm
Fig.1b. Two-dimensional $(\mathrm{C}_{20})_{4\times4}$ structures with the combination of the \textit{open}-{[}2+2{]} and twisted intercluster bonds, configuration B. The binding energy of the fullerenes is $\Delta E=7.08$ eV/$\mathrm{C}_{20}$.

\newpage
\includegraphics[width=\hsize,height=17cm]{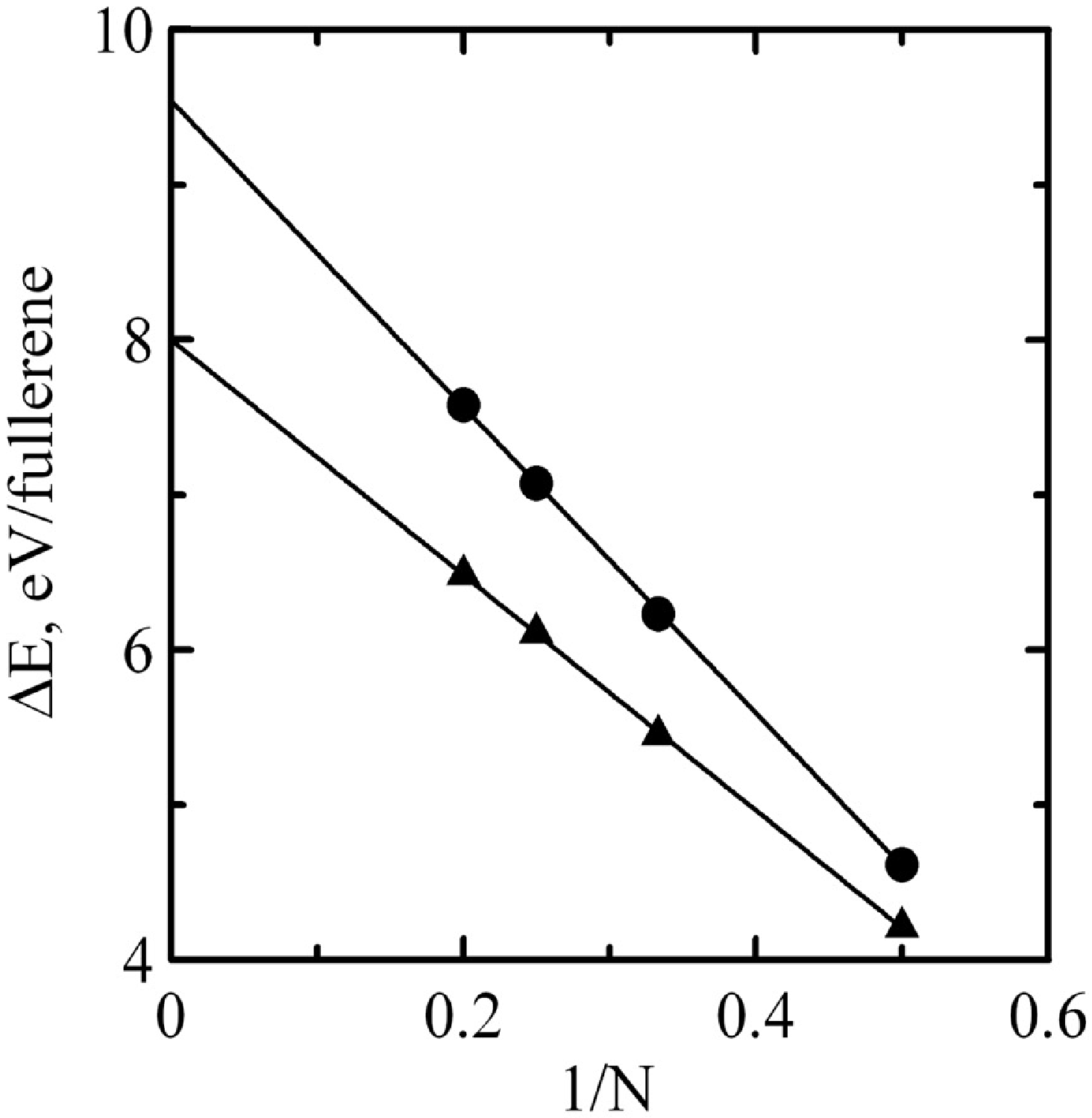}
\vskip 15mm
Fig.2. Binding energies$\Delta E$ of the $\mathrm{C}_{20}$ fullerenes in the two-dimensional $(\mathrm{C}_{20})_{\textnormal{M}\times\textnormal{M}}$ structures with various types of intercluster bonds versus $1/M$ for configurations A (triangles) and B (circles). The solid lines are the least-squares approximations.

\newpage
\includegraphics[width=\hsize,height=15cm]{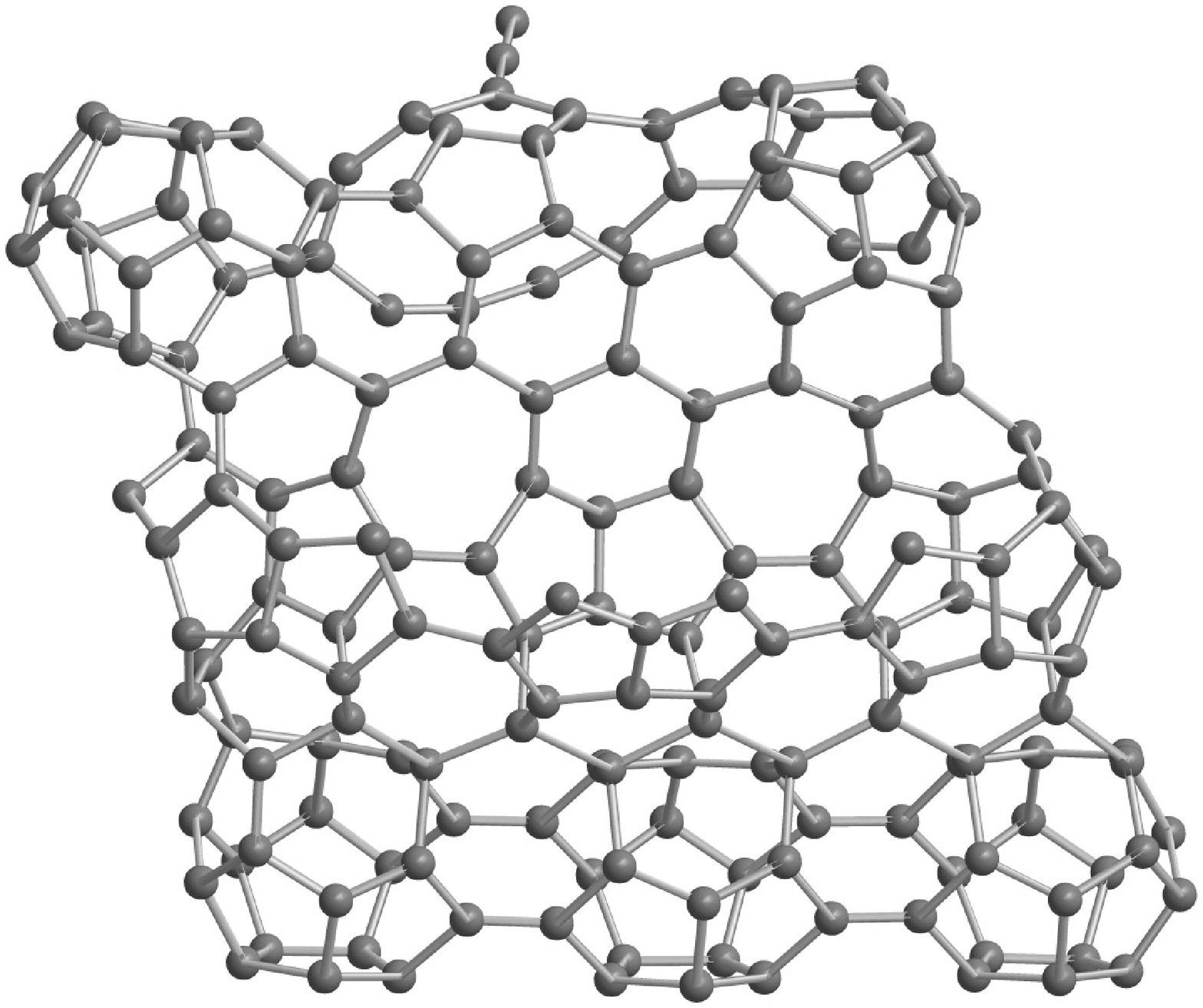}
\vskip 15mm
Fig.3a. Configuration of $(\mathrm{C}_{20})_{3\times3}$ formed after the
decay of one of the $\mathrm{C}_{20}$ fullerenes. The initial temperature
is $T=2500$ K, the lifetime with respect to the decay is $\tau$ = 2.7 ps, and the binding energy (after relaxation) is $\Delta E=5.93$ eV/$\mathrm{C}_{20}$.

\newpage
\includegraphics[width=\hsize,height=15cm]{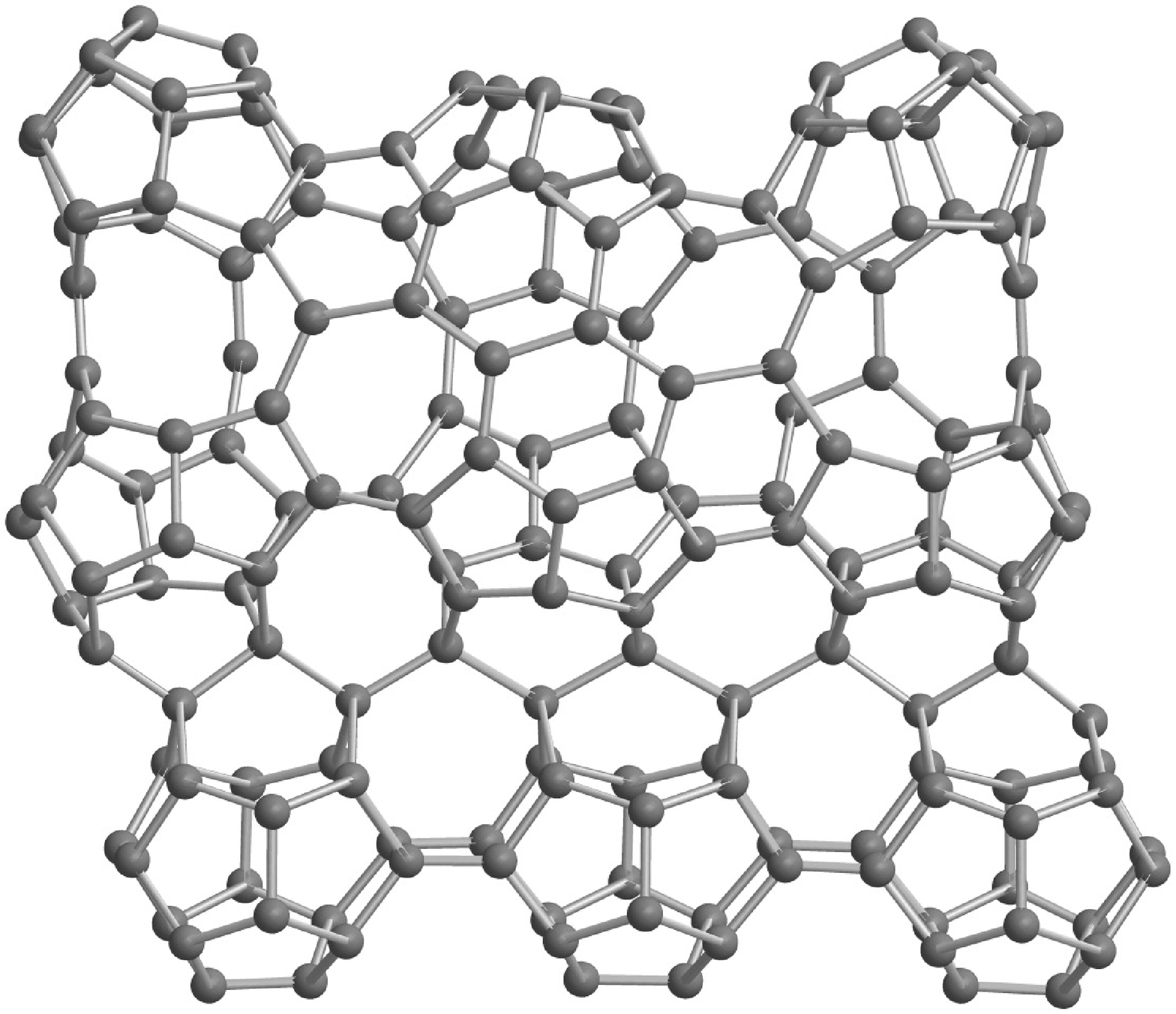}
\vskip 15mm
Fig.3b. Configuration of $(\mathrm{C}_{20})_{3\times3}$ formed after the
coalescence of two $\mathrm{C}_{20}$ fullerenes into the $\mathrm{C}_{40}$ cluster. The initial temperature is $T=2220$ K, the lifetime with respect to the coalescence is $\tau$ = 59.6 ps, and the binding energy (after relaxation) is $\Delta E=7.01$ eV/$\mathrm{C}_{20}$.

\newpage
\includegraphics[width=\hsize,height=16cm]{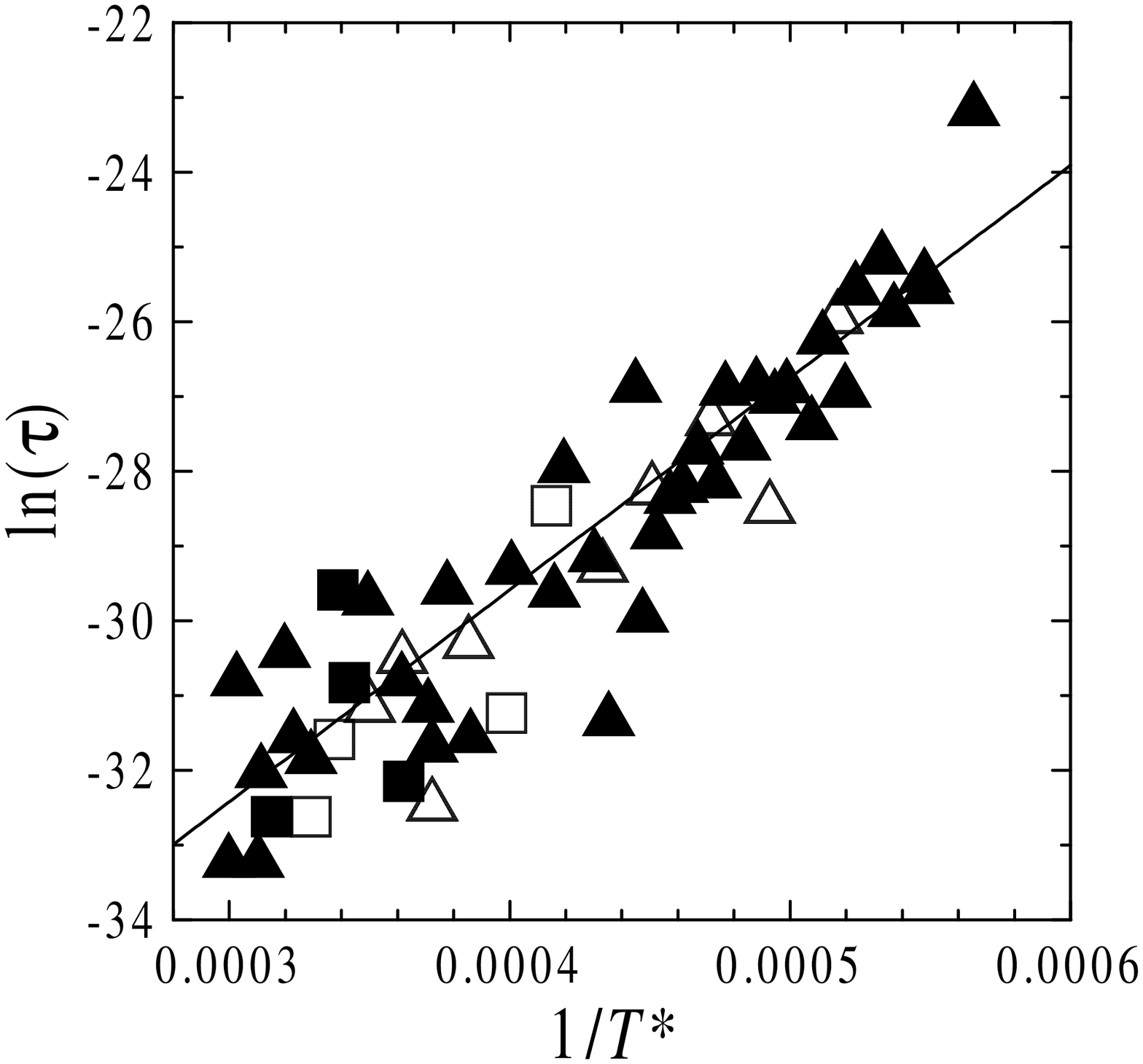}
\vskip 15mm
Fig.4. Logarithm of the lifetime (in seconds) of the
$(\mathrm{C}_{20})_{\textnormal{M}\times\textnormal{M}}$ complexes with $M=$ 2 (closed symbols) and 3 (open symbols) to the instant of the decay of one of the $\mathrm{C}_{20}$ fullerenes (squares) and the coalescence of two $\mathrm{C}_{20}$ fullerenes into the $\mathrm{C}_{40}$ cluster (triangles) versus the inverse initial temperature $T^{*}$ with account for the finite-heat-bath corrections (see the main text). The solid line is the linear approximation for the coalescence processes at $M=2$.

\newpage
\includegraphics[width=\hsize,height=16cm]{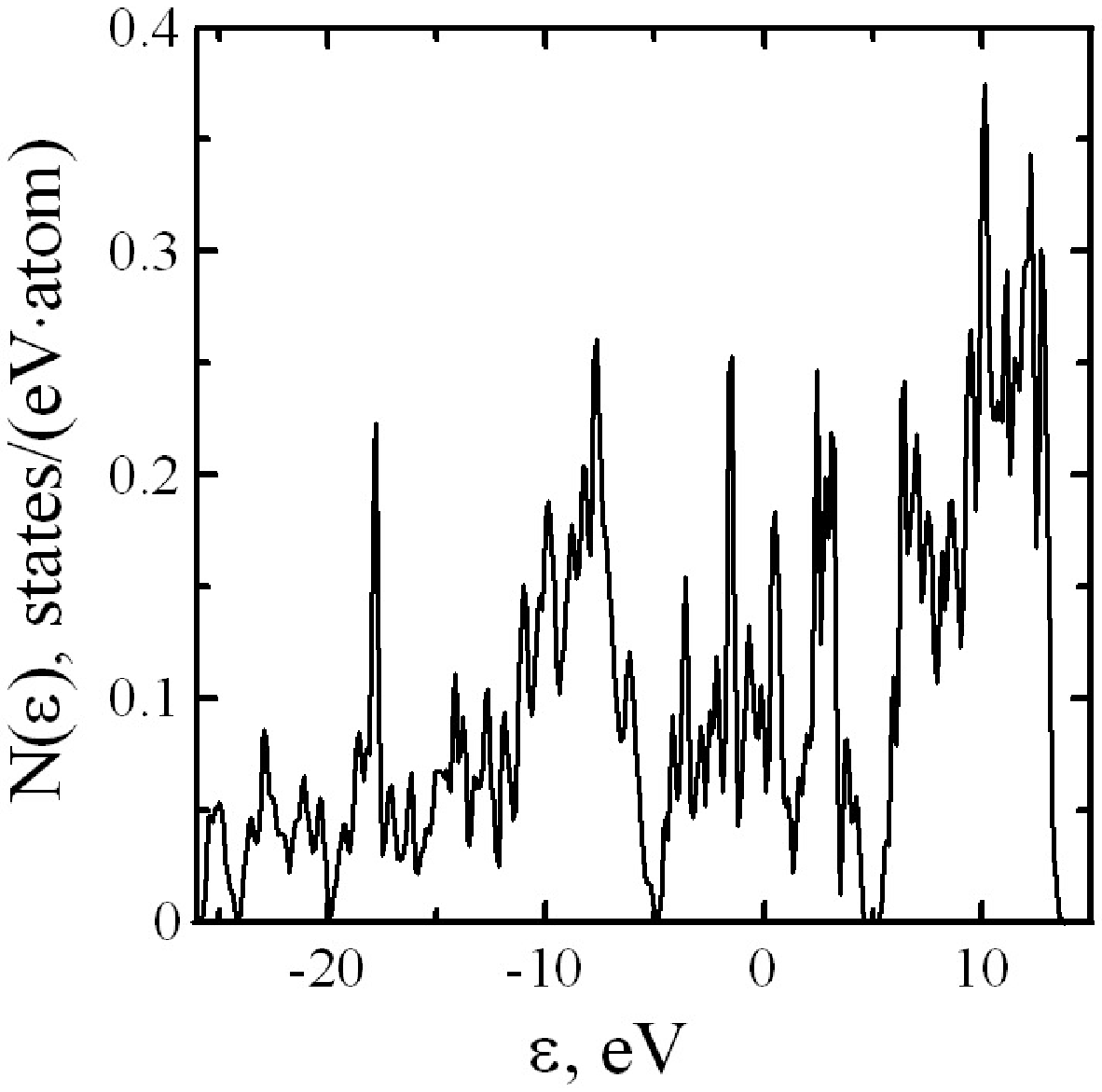}
\vskip 15mm
Fig.5. Density of the single-particle electronic states for the
$(\mathrm{C}_{20})_{4\times4}$ complex with the combination of the \textit{open}-{[}2+2{]} and twisted bonds (see Fig. 1b). The energy is measured from the Fermi level.

\end{document}